\documentclass[twocolumn]{emulateapj}
\usepackage{graphicx}
\usepackage{amssymb}
\usepackage{amsmath}
\usepackage{subfigure}
\usepackage{multirow}
\usepackage{threeparttable}
\usepackage{array}
\usepackage{natbib}
\usepackage{xcolor}
\usepackage{subfigure}
\usepackage{hyperref}

\hypersetup{
    colorlinks=true,
    linkcolor=red,
    filecolor=magenta,
    urlcolor=cyan,
}

\def\cmsq{~\rm cm^{-2}}
\def\cc{~\rm cm^{-3}}

\defcitealias{Henley:2012aa}{HS12}
\defcitealias{Miller:2015aa}{MB15}
\defcitealias{Li:2017aa}{LB17}

\begin{document}
\title{An X-ray and SZ bright diffuse source toward M31: a Local Hot Bridge}
\author{Zhijie Qu$^1$, Rui Huang$^{1, 2}$, Joel N. Bregman$^1$, Jiang-Tao Li$^1$}
\affil{$^1$ Department of Astronomy, University of Michigan, Ann Arbor, MI 48109, USA}
\affil{$^2$ Department of Astronomy, Tsinghua University, Beijing 100084, China}
\email{quzhijie@umich.edu}

\begin{abstract}
We report a large-scale ($r\approx 20^\circ$) X-ray and Sunyaev-Zeldovich (SZ)-bright diffuse enhancement toward M31, which might be a Local Hot Bridge connecting the Milky Way (MW) with M31.
We subtract the Galactic emission from the all-sky \ion{O}{7} and \ion{O}{8} emission line measurement survey, and find that the emission of these two ions is enhanced within $r\approx20^\circ$ around M31.
The mean emission enhancements are $5.6\pm 1.3$ L.U., and $2.8\pm0.6$ L.U. for \ion{O}{7} and \ion{O}{8}, respectively ($>4\sigma$ for both ions).
We also extract the SZ signal around M31, which suggests a surface brightness $y$ of $2-4\times10^{-7}$, an enhancement $>2.5\sigma$ (and a best fit of $5.9\sigma$).
These three measurements trace the hot gas with a temperature $\log~T({\rm K})> 6$, showing similar plateau shapes (flat within $\approx15^\circ$, and zero beyond $\approx30^\circ$).
A single-phase assumption leads to a temperature of $\log~T({\rm K})=6.34\pm0.03$, which is determined by the \ion{O}{7}/\ion{O}{8} line ratio.
Combining X-ray and SZ measurements, we suggest that this feature is unlikely to be the hot halo around M31 (too massive) or in the MW (too high pressure and X-ray bright).
The plateau shape may be explained by a cylinder connecting the MW and M31 (the Local Hot Bridge).
We constrain its length to be about 400 kpc, with a radius of 120 kpc, a density of $\approx 2\times10^{-4}-10^{-3} \cc$, and a metallicity of $0.02-0.1~ Z_\odot$.
The baryon mass is $\gtrsim10^{11}~M_\odot$, and the oxygen mass is about $\gtrsim10^8~M_\odot$, which contribute to the baryon or metal budget of the Local Group.
\end{abstract}
\maketitle

\section{Introduction}
As the best-studied galaxy, the Milky Way (MW) is found to suffer from the missing baryon problem.
In the past decade, multi-wavelength observations revealed that the multi-phase medium within the virial radius of the MW could only account for $\approx 10^{11}~M_\odot$ of baryons \citep{Anderson:2010aa, Gupta:2012aa, Miller:2015aa, Zheng:2019aa, Qu:2020aa}.
Considering the MW halo mass of $1-2\times 10^{12}~M_\odot$ \citep{Xue:2008aa}, about half of the expected baryons are still missing from observations (adopting the cosmic baryonic fraction of 0.158; \citealt{Planck-Collaboration:2016aa}).
One possible solution to this missing baryon problem is that baryons are beyond the virialized halo of the MW.

It is well known that the MW is embedded in the local group (LG), which is dominated by two member galaxies with similar masses: the MW and the Andromeda galaxy (M31; \citealt{Einasto:1982aa}).
The halo mass of the LG is found to be $\log M \approx 12.26-12.83$ and hosts a hot gas-dominated multi-phase medium by matching the local environment (e.g., satellite galaxies) with simulations \citep{Li:2008aa, Nuza:2014aa}.
Observationally, cool-warm clouds ($\log T\approx 4-5$) are detected towards both M31 and anti-M31 directions by detecting the ultraviolet (UV) and \ion{H}{1} high velocity clouds \citep{Bouma:2019aa}.
However, these cool-warm gases only contribute to a small fraction of the total gas mass, and it is still an open question as to the properties of the major gaseous medium associated with the LG.

Observations of X-ray and Sunyaev-Zeldovich (SZ) effect are two effective means to detect the hotter gas than the UV-tracing gas \citep{Bregman:2007aa}.
X-ray ions \ion{O}{7} and \ion{O}{8} are the two most common high ionization state ions in the Universe, and trace gases at $\log T = 5.5$ to 6.8.
The SZ signal is sensitive to all hot gas (electron; $\log T>6$).
Previous studies show that the hot gas covering the entire sky is mainly Galactic rather than from the LG \citep{Bregman:2007ab}, which is determined by the spatial distribution of the \ion{O}{7} absorption equivalent width.

In this paper, we analyze the X-ray data (i.e., \ion{O}{7} and \ion{O}{8} line measurements) and the SZ $y$ signal toward the M31 direction.
We discover a $r \approx 20^\circ$ diffuse hot gas feature toward M31, which is confirmed by both X-ray emission and SZ $y$ signals.
This diffuse hot gas is likely to be a Local Hot Bridge connecting the MW and M31, which accounts for a significant baryonic mass.
The adopted data are mainly from \citet[][hereafter, \citetalias{Henley:2012aa}; \ion{O}{7} and \ion{O}{8} line measurements]{Henley:2012aa} and Qu et al. (2020, in preparation; SZ extraction).
The sample and data reduction are briefly introduced in Section 2.
The origins of this feature and physical implications are discussed in Section 3, where we develop a toy model of the Local Hot Bridge.
We summarize key results in Section 4.

\section{Data and Reduction}
In the following analyses, we adopt the distance to M31 of $D_{\rm M31} = 750$ kpc \citep{Riess:2012aa}, and assume that the projected center of the diffuse hot gas is at M31 ($l$, $b$ = $121.17^\circ$, $-21.57^\circ$).
Although the real center is unknown, the commonly used barycenter of LG ($l$, $b$ = $147^\circ$, $-25^\circ$; \citealt{Einasto:1982aa}) is not favored by both X-ray line measurements and SZ $y$ signals.

\newpage


\begin{figure*}
\vspace{-0.3cm}
\begin{center}
\includegraphics[width=0.44\textwidth]{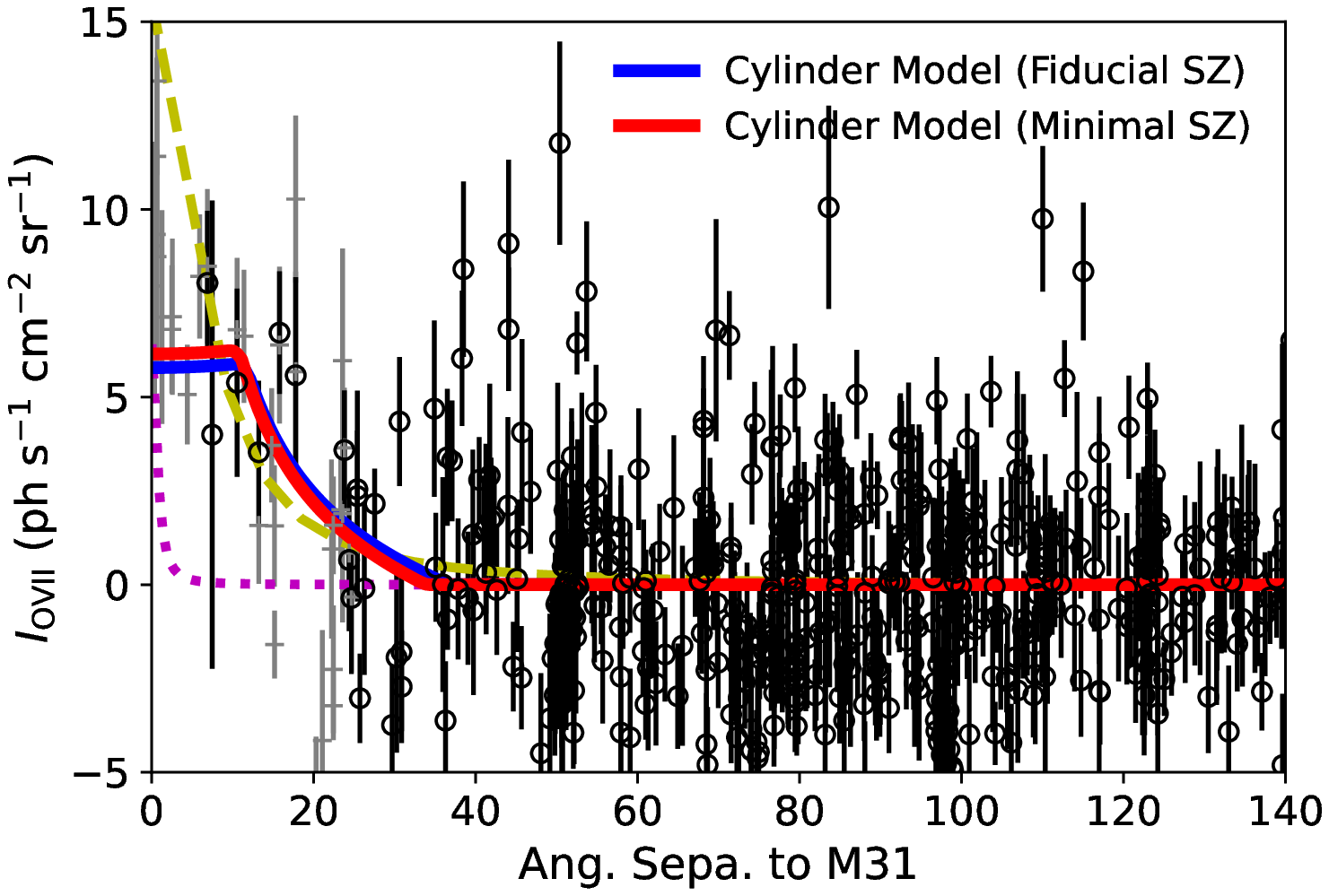}
\includegraphics[width=0.44\textwidth]{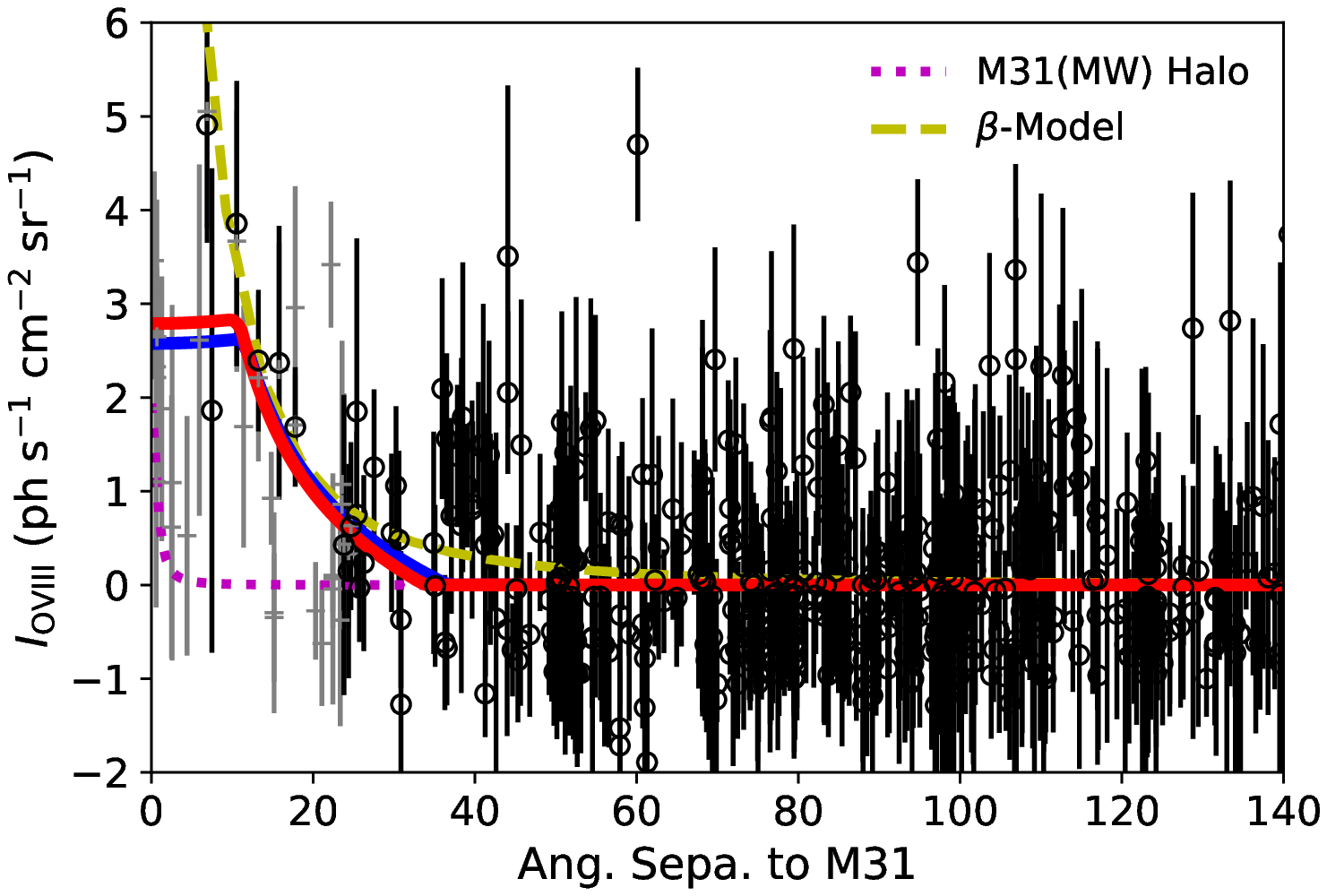}
\includegraphics[width=0.44\textwidth]{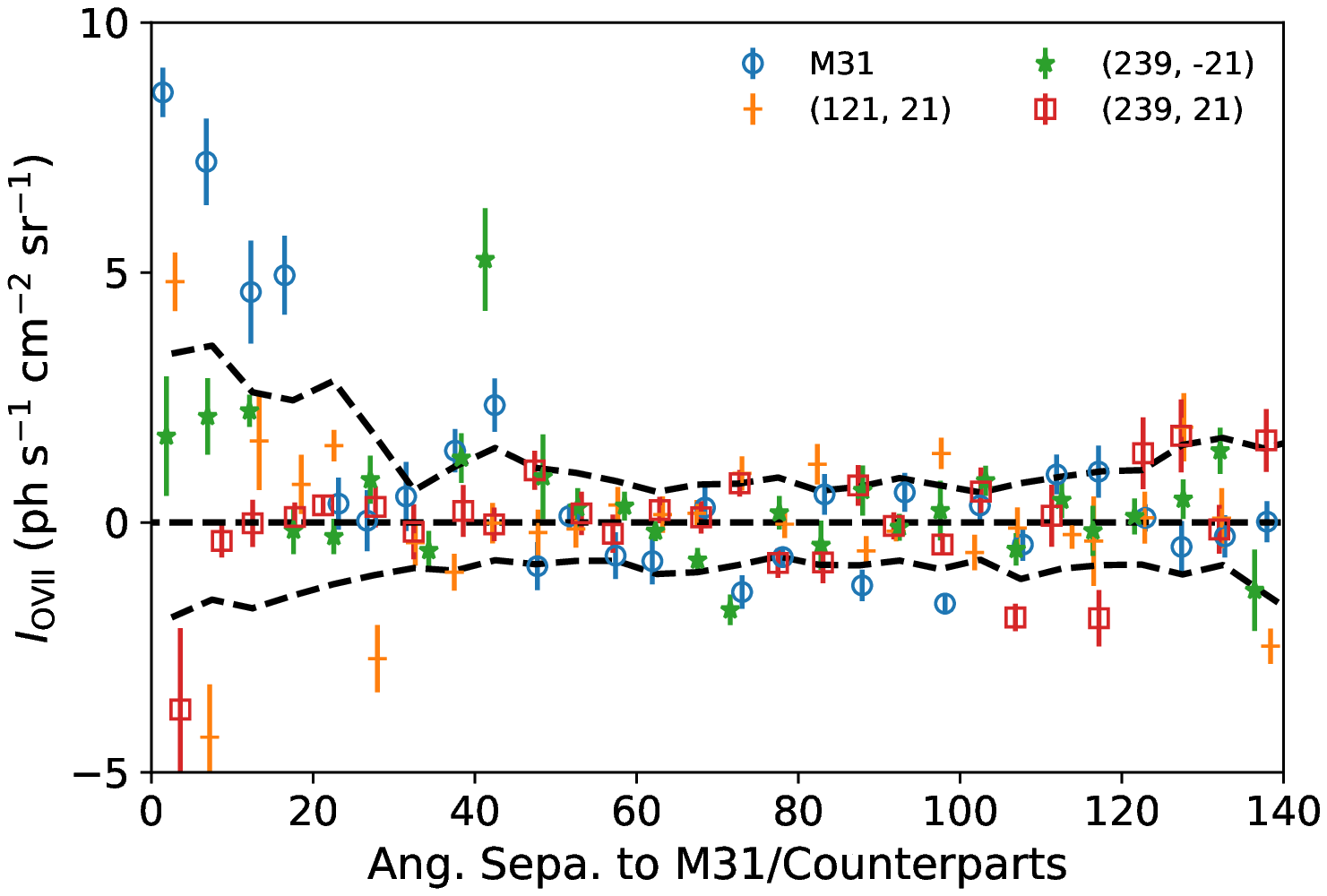}
\includegraphics[width=0.44\textwidth]{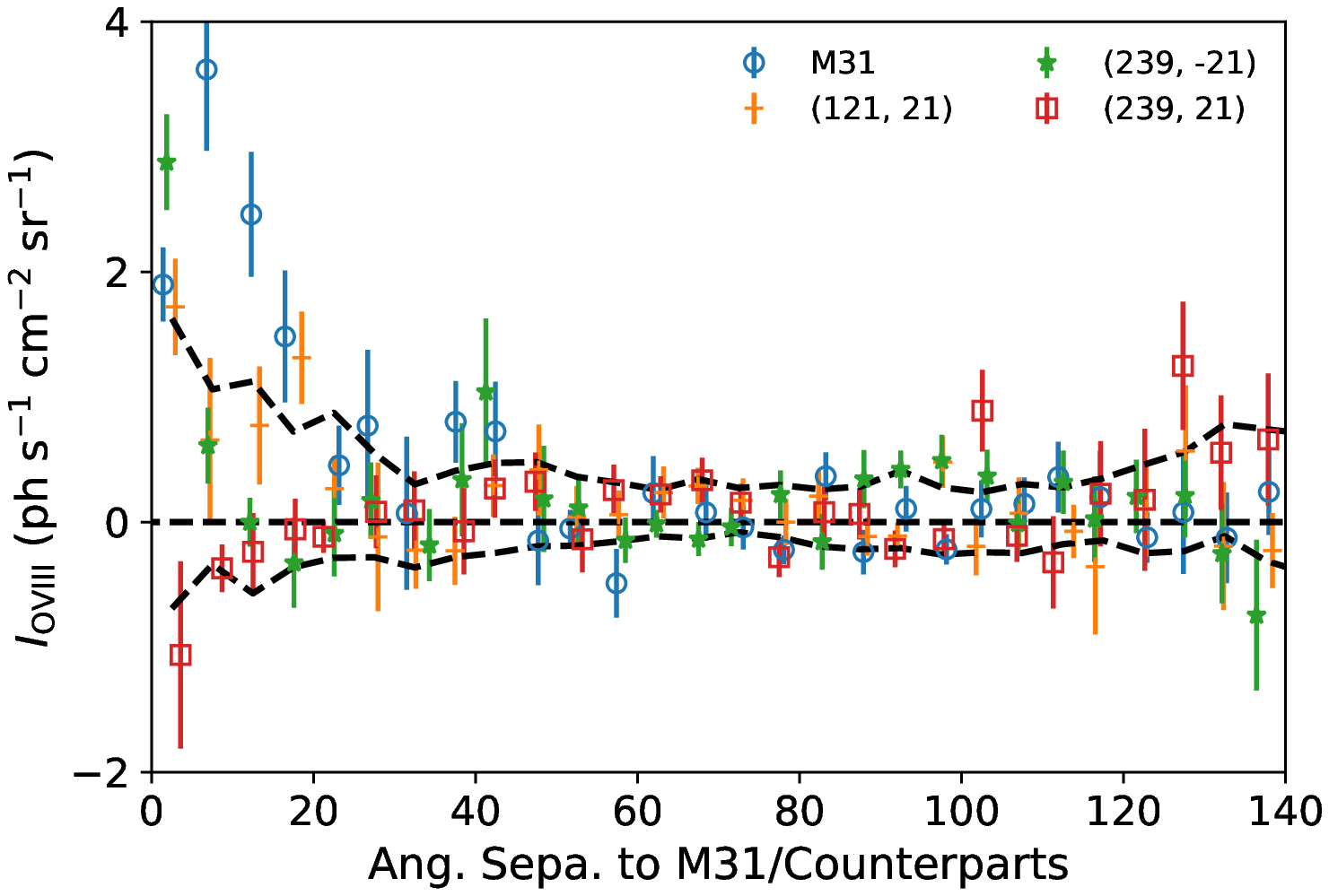}
\includegraphics[width=0.44\textwidth]{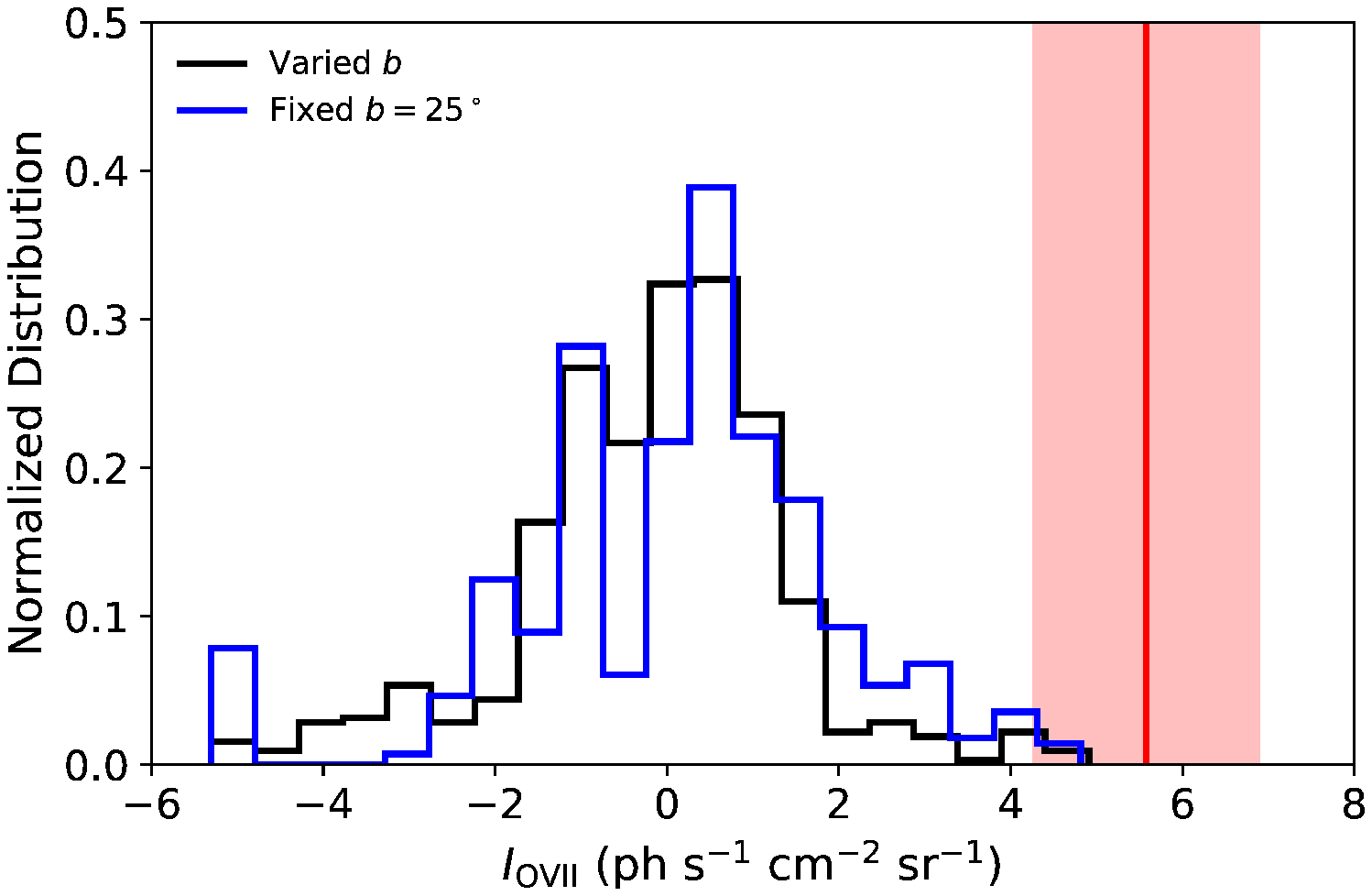}
\includegraphics[width=0.44\textwidth]{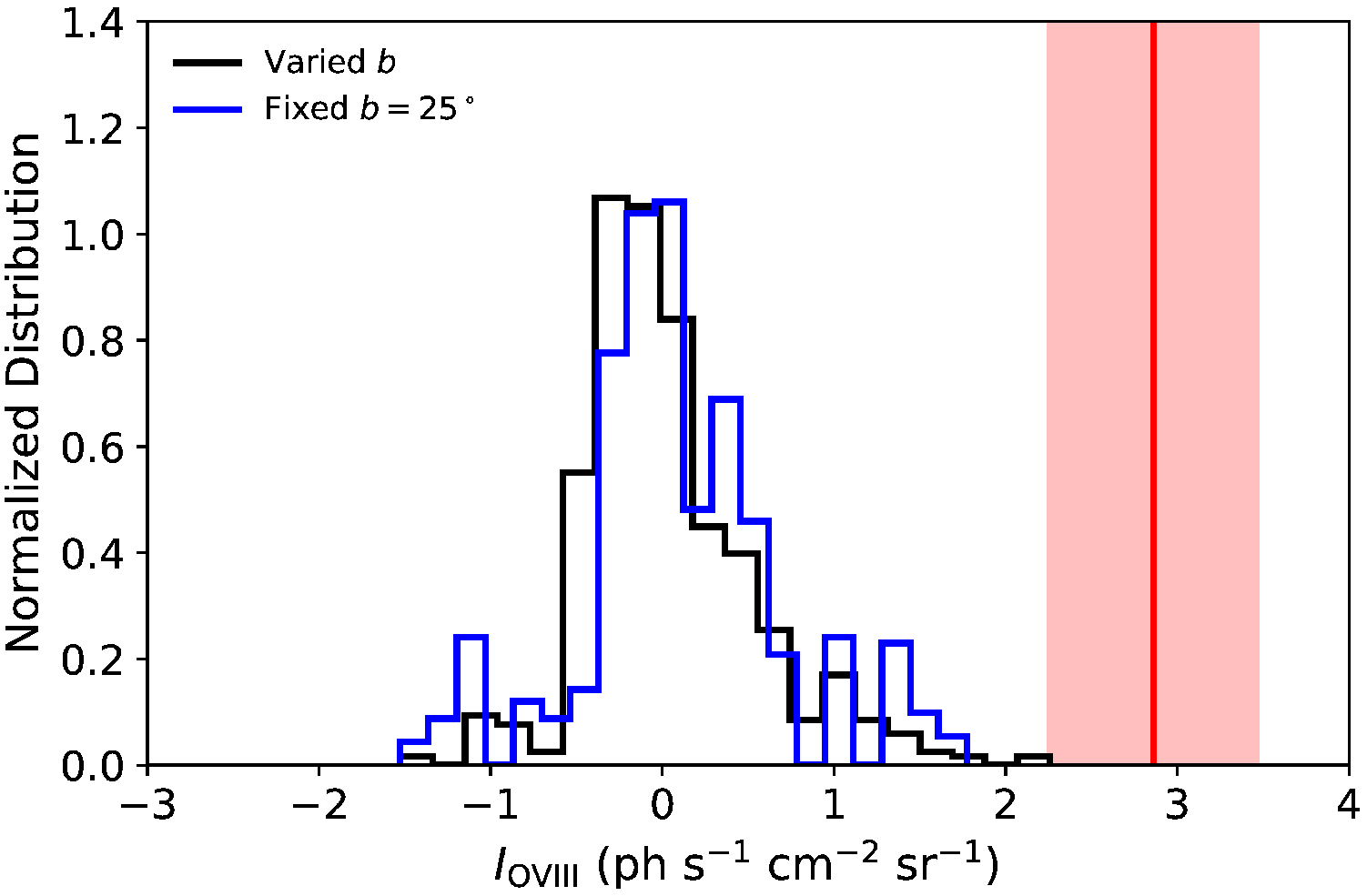}
\end{center}
\vspace{-0.3cm}
\caption{The upper two panels show the X-ray emission line measurements (corrected for the MW model) for \ion{O}{7} (left) and \ion{O}{8} (right).
The black circles are adopted from \citetalias{Miller:2015aa}, while the grey crosses are from \citetalias{Henley:2012aa} (the 1868 sight line sample), which may have more contamination.
For both samples, we masked out the region within $60^\circ$ around the Galactic center, and the measurements with large uncertainties ($>3$ L.U. for \ion{O}{7} and $>2$ L.U. for \ion{O}{8}).
The yellow dashed lines are a projected $\beta$-model based on the input model from the SZ extraction, which has a core of $8^\circ$.
This model systematically overestimates the strengths of \ion{O}{7}, \ion{O}{8} emission measurements and SZ $y$.
The magenta dotted lines are a projected MW-like halo at the distance of M31.
The blue and red solid lines are the Bridge (cylinder) model connecting the MW and M31 for fiducial SZ (left) and minimal SZ (right), which are shown in the Fig. \ref{sz}.
The middle panels show the radial profile ($5^\circ$ bins) of \ion{O}{7} and \ion{O}{8} Galactic-symmetric regions of M31 ($l = 121.17^\circ$ or $238.83^\circ$, $b=\pm21.57^\circ$).
In these two plots, sight lines have the same weights in the \citetalias{Henley:2012aa} and the \citetalias{Miller:2015aa} samples.
The black dashed lines show the $1\sigma$ uncertainty of the radial profile of random sight lines with fixed $b=\pm 25^\circ$.
The signals toward M31 are higher than the other three regions and random sight lines, which indicates it is not some systematical feature associated with the Galactic disk.
In the lower panels, we compare the M31 \ion{O}{7} and \ion{O}{8} measurements to the strength distribution of random sight lines (within $20^\circ$) over the entire sky (black) and with fixed latitudes at $b=\pm 25^\circ$ (blue).
These tests leads to significances of $3.6\sigma$ ($3.0\sigma$ for fixed $b$) and $5.6\sigma$ ($4.8\sigma$) for \ion{O}{7} and \ion{O}{8} measurements, respectively.}
\label{o78}
\end{figure*}

\subsection{The \ion{O}{7} and \ion{O}{8} Emission Measurements}

The adopted \ion{O}{7} and \ion{O}{8} emission line measurements are originally extracted by \citetalias{Henley:2012aa}.
Here, we only briefly describe the criteria for the subset of data used in our study and refer readers to the original paper for the construction of the sample.
Using {\it XMM-Newton} archival data, \citetalias{Henley:2012aa} selected all observations with good time longer than 5 ks (not affected by Solar flares).
They constructed two samples -- a Solar wind charge exchange (SWCX) clean sample of 1868 sight lines (determined by the Solar wind proton flux); and a low extra-galactic emission sample of 1003 sight lines with an additional constraint on the X-ray flux at $2-5$ keV.
The SWCX could introduce non-astrophysical \ion{O}{7} and \ion{O}{8} emissions, which is problematic when extracting the all-sky diffuse emission.
Therefore, a low Solar wind proton flux is crucial to have a clean sample with low SWCX contamination.
The X-ray flux at $2-5$ keV is mainly a criterion to constrain the contamination due to background AGNs.
Using the SWCX-clean and low-background sample, \citet[][hereafter \citetalias{Miller:2015aa}]{Miller:2015aa} applied an additional filter, which cross-matches the {\it XMM-Newton} field (field of view of $ 0.5^\circ$) with known strong X-ray sources (e.g., {\it ROSAT} catalogs and galaxy clusters; \citealt{Voges:1999aa, Piffaretti:2011aa}) to lower possible contamination.
This additional filtering leads to a sub-sample of 649 sight lines.

These 649 sight lines only have 9 sight lines in the $r=25^\circ$ circle around M31.
To use more observations, we also include another 25 sight lines from the original sample (1868 sight lines) of \citetalias{Henley:2012aa}, which are within $r=25^\circ$ around M31, and have small uncertainties ($< 3$ L.U. for \ion{O}{7} and $< 2$ L.U. for \ion{O}{8}; L.U. has units of $\rm ph^{-1}~ cm^{-2}~s^{-1}~sr^{-1}$).
The low uncertainty criterion is also applied to the \citetalias{Miller:2015aa} sample.
We note that 7/25 sight lines are in the M31 disk, which might affect the \ion{O}{7} and \ion{O}{8} extractions because of the thermal component in M31 disk.
These additional sight lines from \citetalias{Henley:2012aa} may have larger contamination compared to the subset used in \citetalias{Miller:2015aa}. 
Therefore, in the following modeling, we will lower the significance of these additional sight lines.

We examine the emission around M31 by subtracting the MW contribution, since the all-sky \ion{O}{7} and \ion{O}{8} emission are dominated by the Galactic hot halo \citep{Henley:2013aa}.
Here, we adopt the \citet{Li:2017aa} model (model No. 9 in Table 1; hereafter \citetalias{Li:2017aa}; Fig. 5 in \citetalias{Li:2017aa}), which considered a $\beta$-model $n(r) = n_0 (1+(r/r_{\rm c})^2)^{-3/2 \beta}$ and an exponential disk ($n(r_{\rm XY}, z) = \exp(-r_{\rm XY}/r_0-z/z_0)$) with radiative transfer.
For this Galactic model, we adopt the same assumptions as \citetalias{Li:2017aa} to correct the hydrogen absorption and the contribution due to the Local Bubble (LB).

The \ion{O}{7} and \ion{O}{8} line measurement residuals show a north-south asymmetry.
The \ion{O}{7} emission is systematically higher in the northern hemisphere than the southern hemisphere, while the \ion{O}{8} measurements show the opposite trend.
The difference between the two hemispheres is about $10-20\%$.
To better model the MW emission (649 sight line sample), we use two normalization factors to reduce the median values of the residuals to zero for the northern and southern hemispheres, respectively.

In Fig. \ref{o78}, we show the residuals projected around M31.
It is clear that both \ion{O}{7} and \ion{O}{8} emission measurements show enhancements in addition to the Galactic emission.
This enhancement shows a plateau shape within $\theta_0 \approx 15^\circ$ of M31, and decays to the zero beyond $\theta_1 \approx 30^\circ$.
The extra sight lines from \citetalias{Henley:2012aa} are consistent with the \citetalias{Miller:2015aa} sample for the \ion{O}{7}.
The \ion{O}{8} emission measurements shows 3 additional sight lines from \citetalias{Henley:2012aa} are slightly lower ($\approx1$ L.U.) than the plateau of the \citetalias{Miller:2015aa} sample within $10^\circ$.

The X-ray emitting region around M31 has an angular diameter of $40^\circ $.
Within $r = 20^\circ$, the \ion{O}{7} enhancement (Galactic emission subtracted) has a mean value of $5.6\pm1.3$ L.U., which is about the same level as the all-sky Galactic \ion{O}{7} emission ($\approx 5-6$ L.U. for the \citetalias{Miller:2015aa} sample).
The \ion{O}{8} enhancement is about $2.8\pm 0.6$ L.U. within $r =20^\circ$, which is higher than the Galactic \ion{O}{8} emission ($\approx 1.3$ L.U.).
Using the additional sight lines from \citetalias{Henley:2012aa}, the mean values are $5.5\pm 0.5$ L.U. and $1.8 \pm 0.3$ L.U. for \ion{O}{7} and \ion{O}{8} (excluding the central 2$^\circ$ to avoid the M31 halo or disk contribution).
The final detection significance is given in Section 3.2 by a Markov chain Monte Carlo (MCMC) model, which is slightly lower ($4.8 \sigma$ and $4.5 \sigma$) because we lower the weights of additional sight lines in the \citetalias{Henley:2012aa} sample.

There are two caveats for this extraction, because the \citetalias{Li:2017aa} model only models the large scale variation of the Galactic emission, and is dominated by the hot halo of a $\beta$-model.
First, some X-ray studies suggest that the Galactic emission is dominated by a disk component \citep{Nakashima:2018aa, Kaaret:2020aa}.
If the disk component is not correctly accounted for in the \citetalias{Li:2017aa} model, it is possible that a variation over Galactic latitudes (higher at low latitudes) leads to the observed feature around M31 because of the low latitude of M31.
In Fig. \ref{o78}, we plot the radial profiles of both \ion{O}{7} and \ion{O}{8} measurements for the Galactic-symmetrical regions of M31 ($l = 121.17^\circ$ or $238.83^\circ$, and $b = \pm 21.57$).
The signal toward M31 is higher than other directions, which disfavor the possibility that the observed feature around M31 is due to unaccounted large-scale variations (i.e., the disk component).

To further investigate this possibility, we also extract 300 random sight lines over all Galactic longitudes, but with limited Galactic latitudes ($b=\pm 25^\circ$) to represent the disk variation at similar latitudes of M31.
The $1\sigma$ radial profile uncertainty is plotted in the middle panels of Fig. \ref{o78}.
It is clear that only the M31 direction shows a significant enhancement within $20^\circ$ away from M31.
We also note that the mean residuals of \ion{O}{7} and \ion{O}{8} emission are slightly positive around M31 ($0.1-0.2$ L. U.), which may be evidence for the disk component enhancement.
However, the significance of this enhancement is about $0.5\sigma$ and $0.8 \sigma$ for \ion{O}{7} and \ion{O}{8}, respectively, which is likely to be random variation.
Also, even if this enhancement is real, it only affects the measurement of the X-ray enhancement around M31 by $2-10\%$, and should not affect the detection of the M31 enhancement.
However, it may influence the derivation of the mass and the metallicity (see discussions in Section 3.4).

Second, the auto-correlation suggests there are remained features $<20^\circ$ in the residuals of \ion{O}{7} and \ion{O}{8} measurements.
We simulate random sight lines to test whether the feature around M31 is due to the random variation of all-sky Galactic emission rather than a disk variation.
We extract 1000 random sight lines over the entire sky, and calculate the median of residuals within $20^\circ$ for each sight line.
We mask out the $30^\circ$ region around M31 to avoid a contribution from the M31 feature to null tests.
None of these random sight lines has a similar strength of the feature seen toward M31. 
Based on this test (Fig. \ref{o78}), the significance is $3.6\sigma$ and $5.6\sigma$ for \ion{O}{7} and \ion{O}{8}, respectively.
For the simulation of limited latitudes ($b=\pm 25^\circ$), the extracted significance is reduced to $3.0\sigma$ and $4.8\sigma$ for \ion{O}{7} and \ion{O}{8}.
We note that \ion{O}{7} is more affected by features in the disk (e.g., supernova remnants), leading to larger residuals and a somewhat lower significance.

\subsection{The SZ $y$ Extraction}
The adopted SZ data in this work will be described in Qu et al. (2020, in preparation) as a part of the all-sky large scale SZ signal.
Here, we briefly discuss the data reduction.
We combine the nine-year {\it WMAP} \citep{Bennett:2013aa} and the Planck data release 3 (PR3) single frequency maps \citep{Planck-Collaboration:2018ab} to extract the SZ signal.
A low-pass filter is applied to extract large-scale features (FWHM $> 5^\circ$).
To avoid the dust contamination, we masked out $40\%$ of the highest intensity dust region around the sky (determined in the Planck 353 GHz map), and the PCCS catalog for point sources \citep{Planck-Collaboration:2016ac}.
We also exclude the region around the ecliptic plane ($\pm 10^\circ$), because the Zodiacal dust contribution is not fully removed in the PR3 maps, showing significant zodiacal contamination of the SZ $y$ (Qu et al. 2020 in preparation).
After these exclusions, $22\%$ of the sky remains.
Toward M31 there are useful SZ signals from the half of $b \lesssim -20^\circ$, mainly due to Galactic dust exclusion regions.

\begin{figure*}
\vspace{-0.3cm}
\begin{center}
\includegraphics[width=0.44\textwidth]{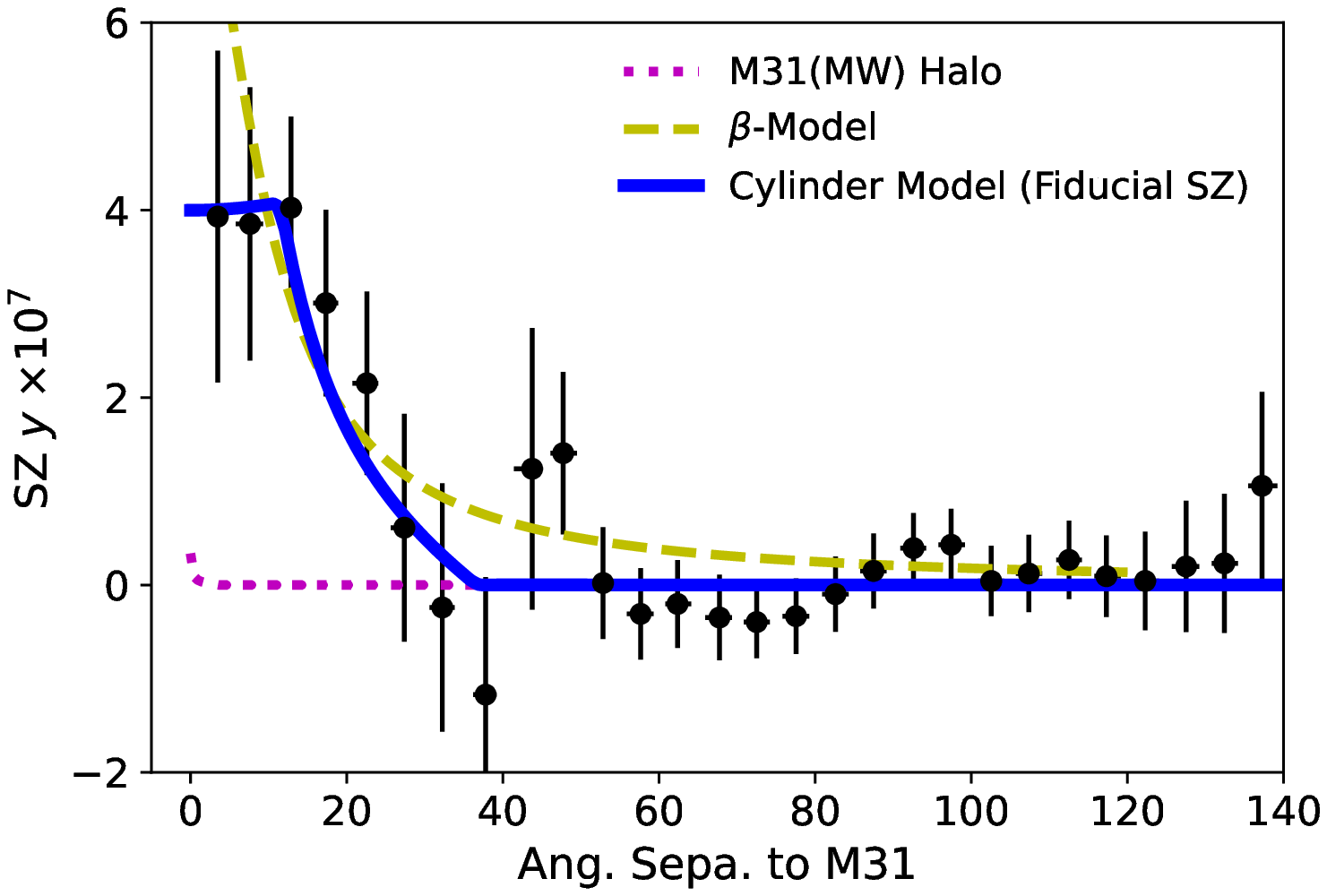}
\includegraphics[width=0.44\textwidth]{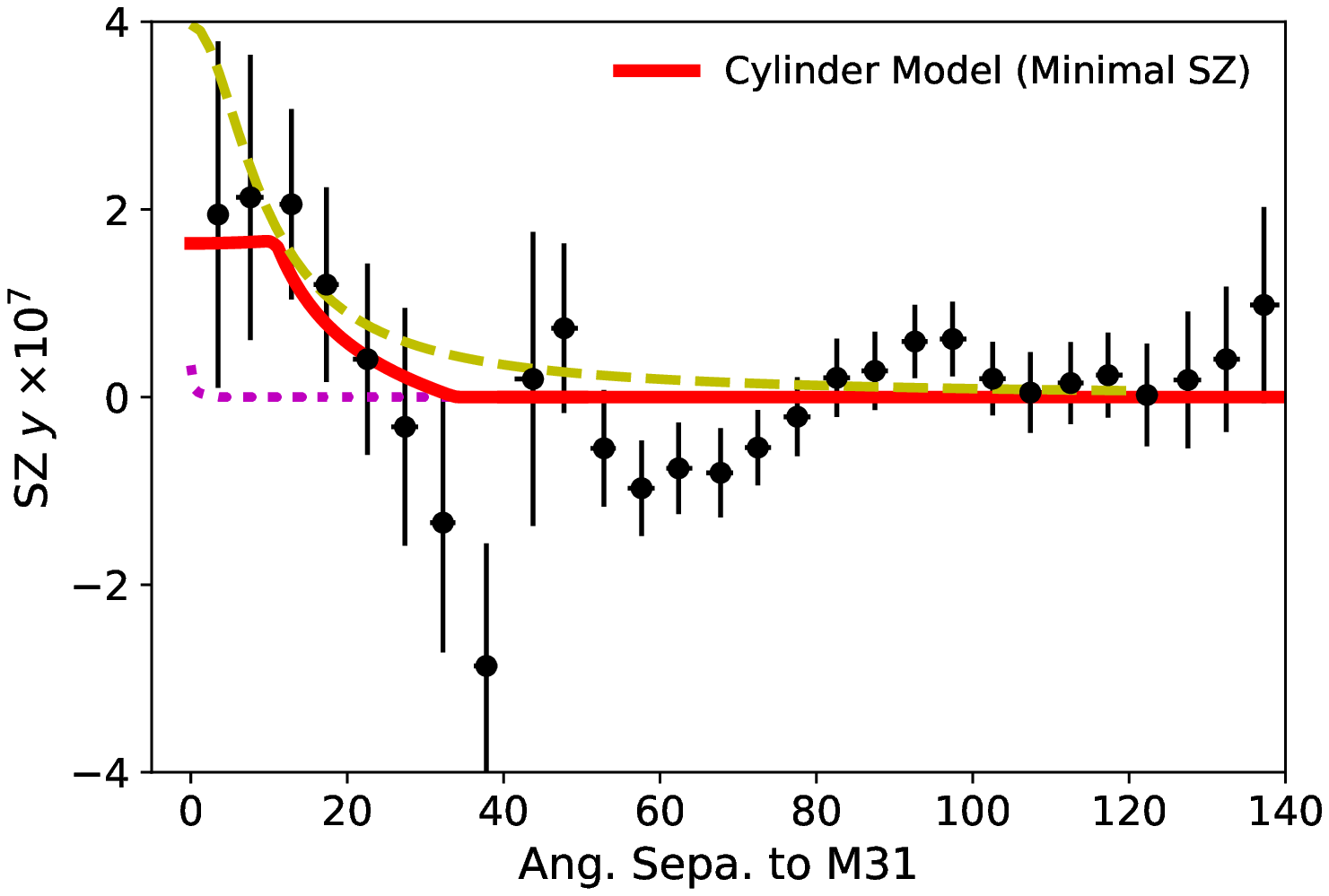}
\includegraphics[width=0.44\textwidth]{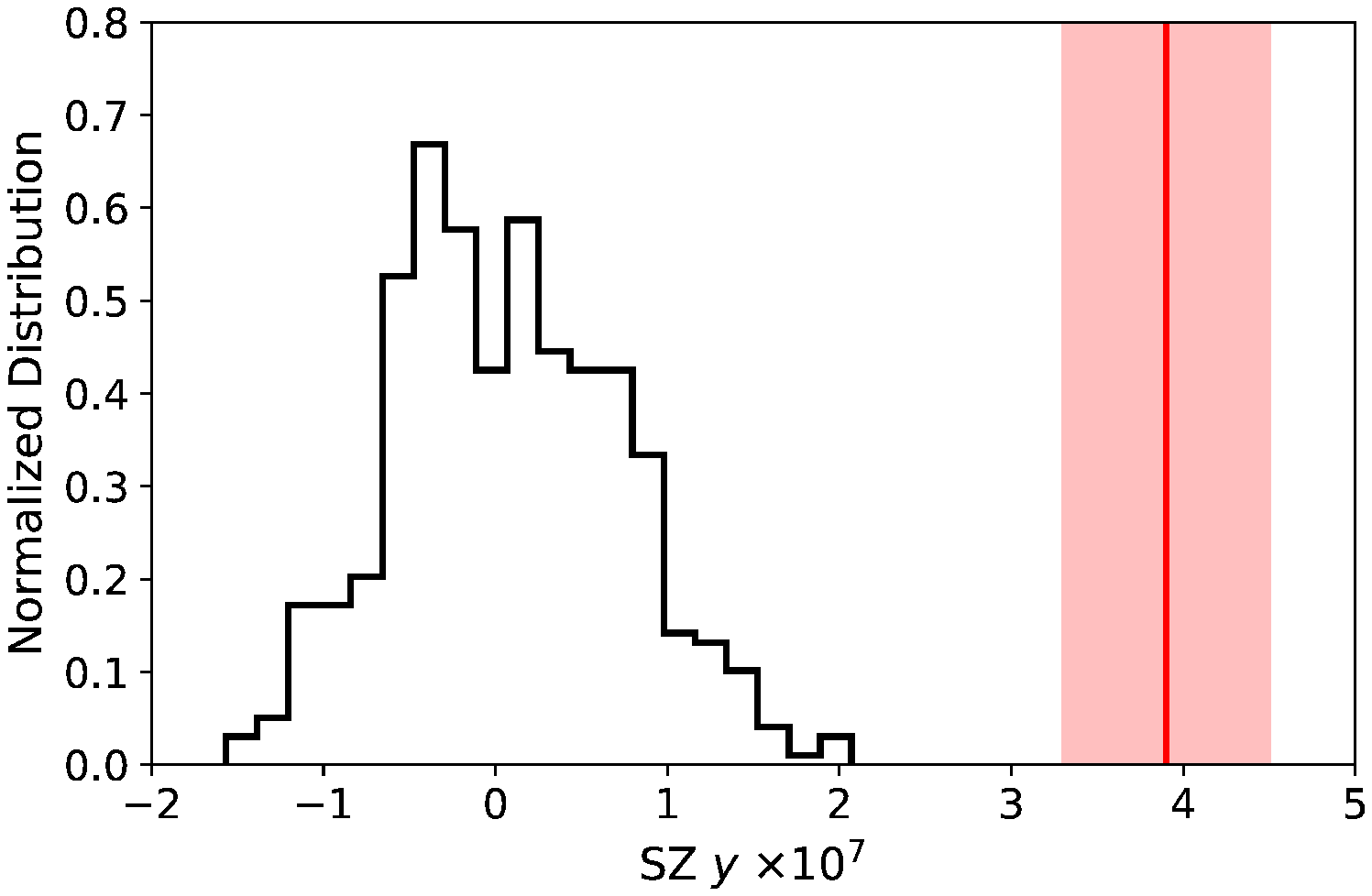}
\includegraphics[width=0.44\textwidth]{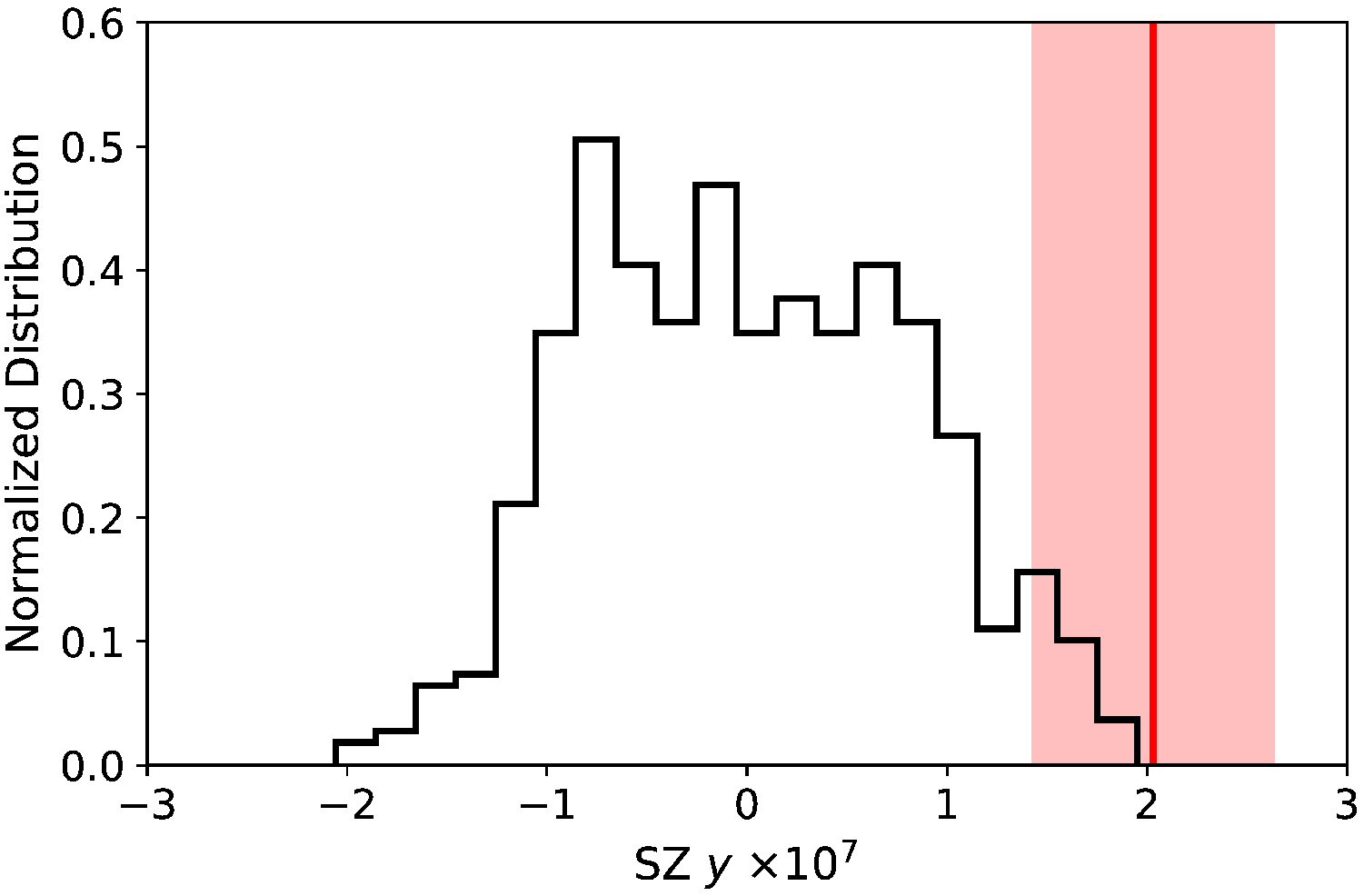}
\end{center}
\vspace{-0.3cm}
\caption{The upper two panels show the fiducial SZ extraction (left) and the minimal SZ extraction (right), while different models have the same colors as Fig. \ref{o78}.
For the SZ extraction, the data points are binned in 5$^\circ$ bins.
The fiducial SZ extraction, we include all four components (i.e., the MW, M31, the local Universe, and the cosmic SZ) in the extraction, which may overestimate the SZ strength.
The feature at $45^\circ$ in the fiducial extraction is a random variation with a significance of $1.8\sigma$, which is a result of the small area left after our heavy masking of the Galactic disk, the ecliptic plane, and point sources.
By disabling the model of M31 and the MW, we extract the minimal SZ strength (see the text for details).
The $\beta$-model is scaled down by a factor of 2 in the minimal SZ plot.
The lowers panels show random sight line tests for the two SZ extractions, which leads to significance of $5.9\sigma$ and $2.5\sigma$.}
\label{sz}
\end{figure*}

We use the internal linear combination (ILC) method to extract large scale features.
The standard ILC method minimizes the variance of the extracted SZ map, which could introduce bias reducing the SZ signal strength (more details in \citealt{Eriksen:2004aa, Delabrouille:2009aa}).
Here, we use the ILC to do the model fitting, which has an input model for the large scale features.
We minimize the variance of the SZ residual maps (i.e., the difference between the extracted SZ map and the input model) instead of the extracted SZ map (the standard ILC).
Therefore, this is a model-dependent extraction of the SZ signal, and we vary the input model to minimize the variance of the residual map.
Then, the SZ extraction and model fitting are performed simultaneously.

Our fiducial input model includes four large scale features: the foreground MW SZ, the feature around M31, the local Universe SZ (e.g., the Virgo cluster), and the cosmic SZ background (i.e., the integration over all redshifts).
The MW foreground SZ is decomposed into two components: a Gaussian disk; and a spherical $\beta$-model halo.
The feature around M31 is modeled as a cored power-law ($y(\theta) = y_0 (1+(\theta/\theta_c)^2)^{-\alpha}$), where $\theta$ is the angular distance from the center of M31, and with a fixed core radius of $\theta_c = 8^\circ$ (100 kpc at M31).
The local Universe SZ is constructed based on the low-$z$ galaxy group and cluster catalog from \citet[][$\log M_{\rm halo} > 13$]{Lim:2017ab}.
For each halo, the total SZ $Y$ is calculated by using the mass-SZ $y$ scaling relationship from \citet[][extrapolated from $\log M=$13.25 to 13]{Pratt:2020aa}, and the universal pressure profile is adopted from \citet{Arnaud:2010aa}.
The cosmic SZ is modeled as a constant over the entire sky since the large scale variations are included in the local Universe SZ component.
Using the MCMC model, we constrain the parameters in these components.
Our fitting results suggested that these four components are significant ($> 5 \sigma$).
In this paper, we focus on the feature around M31, while more results on other components will be discussed in Qu et al. (2020, in preparation).

The final M31 features are extracted from the M31 model along with the total residual (the total model subtracted from the extracted SZ map).
We extract the radial profile in bins of $5^\circ$ in Fig. \ref{sz}, where the input model is also plotted.
A cored power-law model over-predicts the SZ signal for $\theta \lesssim 15^\circ$, but the residual could correct this tendency.
The final extracted SZ shows a significant plateau at $y \approx 4\times 10^{-7}$ within $\approx 15^\circ$.
We refer to this extraction as the fiducial SZ $y$ signal, which is preferred.

The model-dependent extraction of signals may have biases that overestimate the signal, so we test whether one could see similar signals without the input model.
In Fig. \ref{sz}, we also show the case in which only the local Universe and the cosmic SZ components are included in the extraction, which leaves out the MW and M31 components.
This extraction shows a similar plateau shape, but the strength is about half of the fiducial extraction ($y \approx 2\times 10^{-7}$).
However, the background around M31 shows large scale structures, which are corrected in the fiducial extraction.
This extraction leads to the minimal SZ signal, because the ILC method has a bias to reduce the signal \citep{Delabrouille:2009aa}.
The ILC bias is the systematic cancelling of the SZ signal due to empirical correlation between the SZ signal and random noise or astrophysical signals (e.g., the dust, the point sources).
According to \citet{Delabrouille:2009aa}, we estimate the ILC bias for our extraction is about $5\times 10^{-8}$, which is considerable relative to the total SZ signal.

The uncertainty of the SZ extraction has two origins, the model uncertainty obtained from the MCMC model, and the residual variation (including measurement uncertainties and contaminations).
Using the MCMC chain, the model uncertainty is extracted, which is less than $10^{-8}$ around M31, so the final uncertainty is dominated by the residual variation.
We calculate the global standard deviation over the entire sky (except for the region around M31 within $30^\circ$), and scaled it by a $-1/2$ power law with the number of independent spherical harmonic modes in each angular bin.
For the entire sky, there are 1466 independent modes for FWHM $>5^\circ$, and 328 modes left after the masking.
Then, we can use the number of pixels to calculate the equivalent number of modes in each bin, and subsequently, the uncertainty.
This uncertainty leads to a reduced $\chi^2 = 0.81$ for regions $> 40^\circ$ from M31, which are expected to have no features.

We use two means to determine the significance of the SZ extraction.
The direct calculation is the integration of the radial profiles within $20^\circ$, which gives a significance of $6.6\sigma$ and $3.3\sigma$ for fiducial and minimal extractions, respectively.
In another estimation, we simulate 1000 random sight lines over the entire sky, and extract the median SZ within $20^\circ$ around these sight lines.
For these sight lines, we require that there should be more than 1000 pixels (pixel size of $\approx 0.5^\circ$) within the $20^\circ$ region (affected by the mask), since the $20^\circ$ region around M31 has 1244 pixels, and a small number of pixels leads to a larger uncertainty.
Based on the SZ distribution of simulated sight lines (Fig. \ref{sz}), we determine that the median SZ signal around M31 is $5.9 \sigma$ away from the random distribution for the fiducial extraction, and $2.5 \sigma$ for the minimal extraction.
The simulation significances are slightly smaller than the local significance, which indicates that there are still unaccounted features in the sky (e.g., small-scale Galactic features or contamination).

\section{Physical Conditions of the Hot Gas}
\subsection{A $\beta$-Model Halo?}

We rule out this feature to be the M31 hot halo for two reasons.
First, one needs a core radius of 200 kpc to explain the plateau of $15^\circ$, which would be quite unusual for a galaxy group (typical values of tens of kpc).
Second, using the SZ signal, one could estimate the total mass at a given temperature.
For M31 (an MW-like galaxy), the halo temperature is about $2-3\times 10^6$ K.
Then, one could estimate the mass of such a SZ feature to be
\begin{equation}
M_{\rm SZ} \approx 2.5 \times 10^{12} M_\odot \frac{{\rm SZ}y_0}{4\times 10^{-7}}  \frac{2.5\times 10^6}{T({\rm K})} (\frac{D({\rm kpc})}{750})^2,
\end{equation}
where ${\rm SZ}y_0$ is the SZ strength of the plateau, $D$ is the distance of the hot gas, and $T$ is the temperature.
Such a massive hot medium exceeds the cosmic baryonic fraction ($\Omega_{b, 0}/\Omega_{m, 0} = 0.158$; \citealt{Planck-Collaboration:2016aa}).
The halo mass of the local group is $\log M = 12.72$ (12.26 to 12.83; $5-95 \%$; \citealt{Li:2008aa}), and the expected total baryonic mass is about $8.3~(2.9-10.7) \times 10^{11} ~M_\odot$.
A mass of  $2.5 \times 10^{12} M_\odot$ is too large by a factor of $3-10$ to be physically plausible.

If this $\beta$-model halo is between M31 and MW (400 kpc to MW), the required core radius is about 100 kpc, which is larger than generally seen in galaxy groups but not unreasonably so \citep{Mulchaey:2000aa}.
The estimated mass will be about $8.8 \times 10^{11}~M_\odot$.
However, a $\beta = 0.5$ model (typical values of galaxy groups) suggests a long tail to larger angles (Fig. \ref{sz}).
Using current data, the SZ signal does not favor a long tail, while the existence of this long tail cannot be distinguished by the \ion{O}{7} and \ion{O}{8} emission measurements (Fig. \ref{o78}).
A varied $\beta$ extraction leads to $\beta > 1$, which is not found for the $\beta$-model of galaxies or galaxy groups \citep{Osmond:2004aa}.
Therefore, we do not favor this explanation either, but this is not a completely unphysical model.

\subsection{A Galactic Source?}
\begin{figure}
\begin{center}
\includegraphics[width=0.49\textwidth]{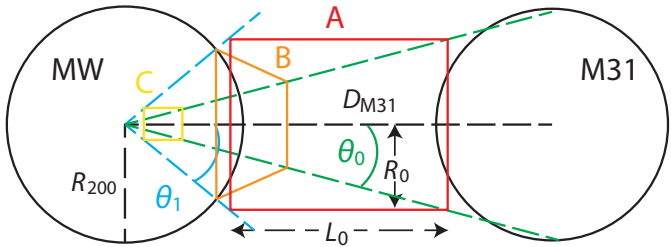}
\end{center}
\caption{ An illustration of the geometry of the Local Bridge. The plateau feature indicates two characteristic angels: the ending angle of the plateau ($\theta_0$) and the angle beyond which the emission is zero ($\theta_1$). Case A is the fitting model described in Section 3.3. Cases B and C are discussed in section 3.2.}
\label{illustration}
\end{figure}

We consider whether the detected diffuse feature belongs to Galactic structures (e.g., Case B and C in Fig. \ref{illustration}).
We derive the scaling relationships between physical parameters with the distance under the observational constraints.
Here, we assume that the hot diffuse structure has a length of $L_0$ and a radius of $R_0$.
Then, the two ratios of $L_0/D$ and $R_0/D$ are determined by $\theta_0$ and $\theta_1$.
The temperature is a constant that is determined by the \ion{O}{7}/\ion{O}{8} ratio.
Two other constraints $nTL_0$ and $Zn^2 \Lambda L_0$ are also constant, determined by the SZ and the \ion{O}{7} or \ion{O}{8} emission.
At a distance of $D_{10}$ (in units of 10 kpc), the scaling relations (with the fiducial SZ value) will be
\begin{eqnarray}
L_0 & \approx & 9.7 D_{10} {\rm~ kpc} \notag\\
R_0 & \approx & 3.2 D_{10}  {\rm~ kpc} \notag\\
n & \approx & 5.0  D_{10}^{-1}  \times 10^{-2} {\rm \cc} \notag \\
Z & \approx & 2.4 D_{10} \times 10^{-3} Z_\odot \notag \\
M & \approx & 4.4 D_{10}^2 \times 10^{8} M_\odot \notag \\
L_{\rm OVII+OVIII} & \approx & 9.2 D_{10}^2 \times 10^{36} {\rm~erg~s^{-1}}  \notag \\
L_{\rm X} & \approx & 3.0 D_{10} \times 10^{40} {\rm~erg~s^{-1}} \notag \\
I_{\rm X} & \approx & 6.2 D_{10}^{-1} \times 10^{37} {\rm~erg~s^{-1}~kpc^{-2}~sr^{-1}}.
\end{eqnarray}
$L_{\rm OVII+OVIII}$ is the total luminosity of \ion{O}{7} and \ion{O}{8}, and $L_{\rm X}$ and $I_{\rm X}$ are the X-ray bolometric luminosity and the X-ray surface brightness.
To convert $L_{\rm OVII+OVIII}$ to $L_{\rm X}$, we adopt the APEC model \citep{Smith:2001aa}.
This conversion factor is proportional to the inverse of the metallicity, because when $Z \lesssim 0.01$, the X-ray emissivity is dominated by  bremsstrahlung emission rather than metal lines.

This feature can not be too close to the Sun (e.g., $D<10$ kpc), or it would have been discovered by all-sky X-ray surveys (e.g., {\it ROSAT}; \citealt{Snowden:1997aa}) because of the high surface brightness ($I_{\rm X}$).
As a comparison, the unabsorbed Galactic X-ray emission is about $3-4\times 10^{35}\rm~ erg~ s^{-1}~ kpc^{-2}~sr^{-1}$.
Also, the mass of the this feature will be larger than $4\times10^6~M_\odot$ at $D>1$ kpc, which is unlikely to be a feature in the disk.

We also suggest that this hot gas structure cannot be in the MW halo ($\sim 100$ kpc) because of the resulting high pressure.
In the MW halo, the typical ambient gas pressure is about $n_{\rm 200} T_{\rm vir} \approx 100\rm ~ K~cm^{-3}$, where $n_{\rm 200} $ is 200 times the critical matter density, and $T_{\rm vir}$ is the virial temperature.
If the detected hot gas structure is about 100 kpc away from the Sun, the hot gas pressure will be about $1.1\times 10^4\rm ~ K~cm^{-3}$, two orders of magnitude greater than expectations.
Therefore, it is very unlikely to be a structure in the MW halo.

\subsection{The Local Hot Bridge Connecting MW and M31 -- A Single-Phase Toy Model}
The plateau of the SZ signal inspires a toy model of a hot bridge connecting the MW and M31.
As suggested by simulations, a hot bridge occurs between the MW and M31 after $z<1$, although these two galaxies have not yet entered each others virial radius \citep{Nuza:2014aa}.
Here, we use the simplest assumption to model the observation -- a single temperature, uniformly distributed medium filling in a cylinder between the MW and M31 (Fig. \ref{illustration}).
The direction of this cylinder is toward M31, and the barycenter is the middle point along the sight line ($375$ kpc).
The length and radius of the cylinder are $L_0$ and $R_0$, respectively.

\begin{table}
\begin{center}
\caption{Properties of the Local Hot Bridge}
\label{phy_params}
\begin{tabular}{lcc}
\hline \hline
 & Fiducial SZ  & Minimal SZ  \\
\hline
$I_{\rm OVII, 0}$ (L.U.) & $5.7_{-1.2}^{+1.3}$ & $6.0_{-1.3}^{+1.4}$\\
$I_{\rm OVIII, 0}$ (L.U.) & $2.5_{-0.5}^{+0.6}$ & $2.8_{-0.6}^{+0.7}$\\
SZ$y_0$ & $3.9\pm0.8\times 10^{-7}$ & $1.62_{-1.0}^{+0.9} \times 10^{-7}$\\
$L_0$ (kpc) & $430\pm 150$ & $420\pm 150$ \\
$D_0$ (kpc) & $120\pm 20$  & $110\pm20$\\
$\log n_{\rm H} ({\rm cm^{-1}})$ & $-2.91_{-0.14}^{+0.17}$ & $-3.29_{-0.45}^{+0.26}$ \\
$\log T({\rm K})$ & $6.35\pm0.03$ & $6.35 \pm 0.03$  \\
$\log M_{\rm hot}(M_\odot)$ & $11.74 \pm 0.11$ & $11.28_{-0.42}^{+0.22}$  \\
$\log M_{\rm oxy}(M_\odot)$ & $7.68 \pm 0.15$ & $8.08_{-0.25}^{+0.46}$ \\
$L_{\rm O VII}$ ($\rm erg ~s^{-1}$) & $8.9_{-2.1}^{+2.2} \times 10^{39}$ & $8.3_{-2.0}^{+2.2} \times 10^{39}$ \\
$L_{\rm O VIII}$ ($\rm erg ~s^{-1}$) & $4.5 \pm 1.0 \times 10^{39}$ & $4.2_{-1.2}^{+1.4} \times 10^{39}$ \\
${L_{\rm X}}^a$ ($\rm erg ~s^{-1}$) & $1.1 \pm 0.3 \times 10^{42}$ & $2.6 \pm 0.8 \times 10^{41}$ \\
$\log Z/Z_\odot$ & $-2.0\pm0.2$ & $-1.2_{-0.4}^{+0.9}$  \\
\hline
\end{tabular}
\end{center}
Note: all parameters in this table is based on the single phase assumption, the correction due to the multi-phase medium is in Section 3.4.2.\\
$^a$ adopting the APEC model to convert the line emissivity to the bolometric luminosity.
\end{table}

It is well known that the CGM is typically multi-phase (\citealt{Tumlinson:2017aa} and references therein).
However, a multi-phase medium model cannot be constrained without direct observations of cool or warm gas associated with the bridge.
In section 3.4, we discuss the limitation (bias) introduced by the single-phase assumption, together with the observational limitations.

The strength of \ion{O}{7} and \ion{O}{8} emission measurements or the SZ $y$ signal is proportional to the path length in the cylinder.
There are two characteristic angles for this bridge model: the opening angle at the M31 side, $\tan\theta_0 = R_0/(D_{\rm M31}/2+L_0/2)$, and the opening angle at the MW side, $\tan\theta_1 = R_0/(D_{\rm M31}/2-L_0/2)$.
Based on these two angles, the path length in the cylinder could be divided into three regimes:
\begin{eqnarray}
L_{\rm Cyl} & = & L_0 / \cos \theta, ~~~0 < \theta < \theta_0, \notag \\
& = & R_0/\sin \theta - D_{\rm min}/\cos \theta ,~~~  \theta_0 \leq \theta < \theta_1, \notag  \\
& = & 0, ~~~\theta_1 \leq \theta,
\end{eqnarray}
where $D_{\rm min}$ is $D_{\rm M31}/2-L_0/2$.

Within the cylinder, the gas is assumed to be well mixed, with the same density, temperature, and metallicity.
For the \ion{O}{7} and \ion{O}{8} emission, we use the AtomDB data set to extract the emissivity at different temperatures \citep{Foster:2012aa}.
A factor of 0.58 is used to correct the solar oxygen abundance difference, $8.5\times 10^{-4}$ in AtomDB \citep{Anders:1989aa}, and we use $4.9\times 10^{-4}$ from \citet{Asplund:2009aa}.

The M31 hot halo may also contribute to the observed X-ray and SZ signals.
Here, we assume that M31 hosts a MW-like hot halo (the \citetalias{Li:2017aa} model) at 750 kpc.
The contributions due to such a hot halo are important within about $1^\circ-2^\circ$ around M31 (Fig. \ref{o78} and Fig. \ref{sz}).
In practice, we subtract the contribution due to this MW-like hot halo from observed signals, before the modeling of the cylindrical hot bridge.

\begin{figure*}
\begin{center}
\includegraphics[width=0.98\textwidth]{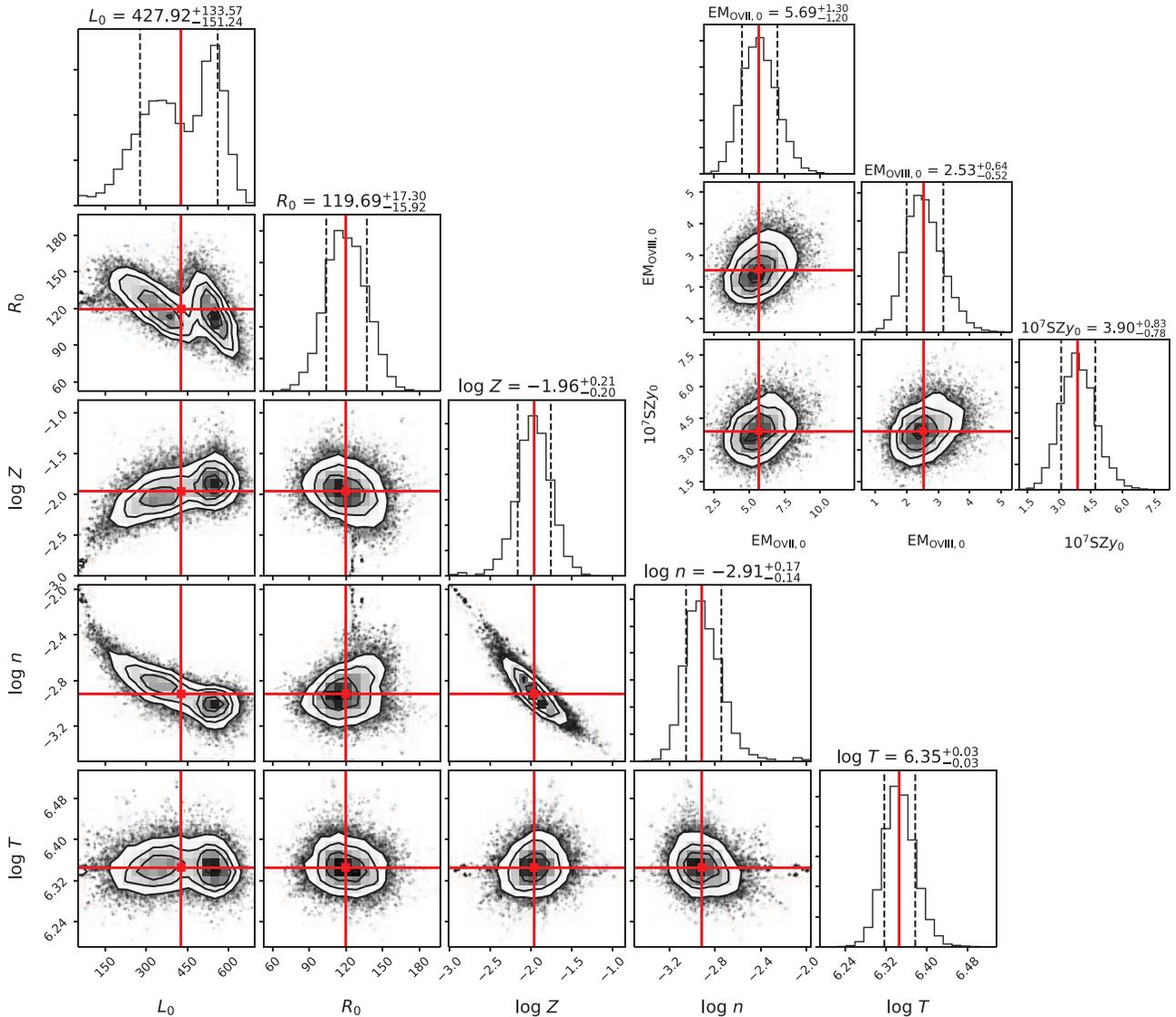}
\end{center}
\caption{The posterior distribution of the toy model for the fiducial SZ extraction. The lower left corner plot is the input physical parameters of the length ($L_0$), the radius ($R_0$) of the cylinder, the number density ($\log n$), the metallicity ($\log Z$), and the temperature ($\log T$). The upper right plot shows the phenomenological parameters derived from the model: the strength of the plateau for \ion{O}{7} and \ion{O}{8} emission measurements and the SZ $y$ strength.}
\label{phy_fiducial}
\end{figure*}

\begin{figure*}
\begin{center}
\includegraphics[width=0.98\textwidth]{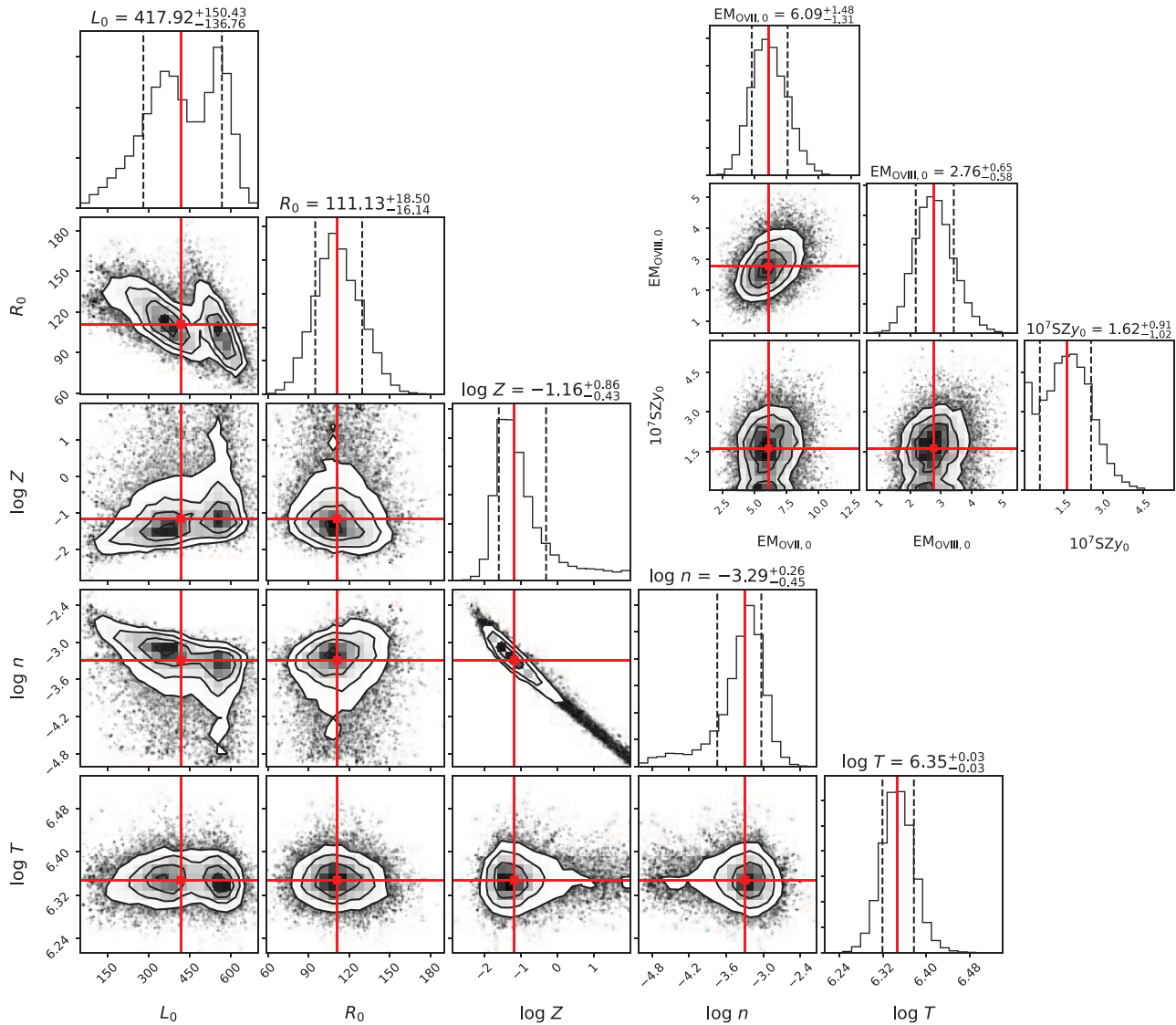}
\end{center}
\caption{Same as Fig. \ref{phy_fiducial}, but for the minimal SZ extraction.}
\label{phy_min}
\end{figure*}

For the fitting, we expect that points within $25^\circ$ significantly contribute to the model constraints.
The \citetalias{Miller:2015aa} sample has 9 sight lines, the SZ radial profile has 5 bins, while the \citetalias{Henley:2012aa} sample has additional 24 sight lines.
As stated in Section 2, the additional sight lines in \citetalias{Henley:2012aa} may suffer from more contamination than \citetalias{Miller:2015aa}, so we lower their weights by a factor of 10 in the fitting.
Then, the \citetalias{Henley:2012aa} sample has about $2-3$ equivalent sight lines, which has slightly lower contributions to the model fitting than the \citetalias{Miller:2015aa} sample and the SZ signal.
We also have an additional uncertainty for \ion{O}{7} of 1.5 L.U., because the \ion{O}{7} is more clumpy, showing small scale variations (\citetalias{Miller:2015aa}, \citetalias{Li:2017aa}).
The total likelihood is
\begin{equation}
\ln p = -\frac{1}{2}(\sum \chi^2_{\rm MB15} + \sum \chi^2_{\rm SZ} + \frac{1}{10}\sum \chi^2_{\rm HS12} ).
\end{equation}
The MCMC model is calculated with {\it emcee} \citep{Foreman-Mackey:2013aa}, and the results are shown in Fig. \ref{phy_fiducial} and Fig. \ref{phy_min} for fiducial SZ and minimal SZ extractions, respectively.
The physical parameters are summarized in Table \ref{phy_params}.

The cylinder model suggests the length of this structure is about $400$ kpc, which is sufficient to connect the dark matter halo of the MW and M31 (250 kpc for each galaxy).
Therefore, it is consistent with the assumption that the observed hot diffuse feature is a Local Hot Bridge connecting the MW and M31.

For the fiducial SZ model, the estimated mass is $\log M (M_\odot) = 11.74 \pm 0.11$, while the minimal SZ model has a mass of $\log M (M_\odot)  = 11.28_{-0.42}^{+0.22}$.
With a halo mass of $\log M = 12.72$, the associated baryon mass is about $\log M = 11.92$ \citep{Li:2008aa, Planck-Collaboration:2018ab}.
Then, the Local Bridge structure contributes about $66\%$ of the baryon mass ($23\%$ in  the minimal model).
As a comparison, we estimate the baryon masses in the MW and M31.
The MW has a stellar mass of $\log M = 10.71\pm0.09$ \citep{Licquia:2015aa}, a hot gas CGM of $\log M = 10.5-11$ (\citealt{Gupta:2012aa}; \citetalias{Miller:2015aa, Li:2017aa}; \citealt{Faerman:2017aa, Faerman:2020aa}), and a warm gas CGM of $\log M \lesssim 10$ \citep{Zheng:2019aa, Qu:2019ab}.
M31 has a stellar mass about twice the MW of $\log M=11$ \citep{Tamm:2012aa}, but a similar halo mass of $\log M \approx 12-12.3$ \citep{Kafle:2018aa}.
The hot component of the M31 CGM is still unknown, but should be comparable to the MW because of the similar halo mass.
The cool-warm CGM in the M31 halo has mass of $\log M = 10.6$ \citep{Lehner:2020aa}.
Therefore, the baryons within the MW and M31 halos account for $\log M = 11.3-11.6$ ($24-48\%$).

We estimate the total oxygen masses are $\log M_{\rm oxy} = 7.68\pm0.15$ and $8.08_{-0.25}^{+0.46}$ for fiducial and minimal SZ models, respectively.
According to \citet{Peeples:2014aa}, the total oxygen generated in a MW-like galaxy is $\log M_{\rm oxy} = 8.7$, and about $20-40\%$ of oxygen ($\log M_{\rm oxy} = 8.0-8.3$) is missing within the virial radius.
Then, the total missing oxygen is about $\log M_{\rm oxy} = 8.3-8.6$, saying M31 is also a MW-like galaxy.
The metals in the Local Bridge also considerably account for the LG missed metals ($10-80\%$).
We note that with a lower SZ strength, the oxygen mass will be higher.

The total \ion{O}{7} and \ion{O}{8} luminosity is about $1.3\pm0.3 \times 10^{40} \rm ~erg ~s^{-1}$.
Adopting the APEC conversion factors \citep{Smith:2001aa}, the bolometric X-ray luminosity is $1.1 \times 10^{42}$ erg s$^{-1}$ (fiducial SZ extraction) and $2.6\times 10^{41}$ erg s$^{-1}$ (minimal SZ extraction), which are comparable to poor galaxy groups \citep{Osmond:2004aa, OSullivan:2014aa}.
A typical galaxy absorption toward M31 of $5\times10^{20} \cmsq$ leads to an observed luminosity of $\approx5-6\times 10^{38}~\rm erg~s^{-1}$ the $0.2-5$ keV band for the fiducial SZ extraction and $\approx1-2\times 10^{38}~\rm erg~s^{-1}$ for the minimal SZ extraction.

\subsection{Caveats and Preferred Parameters}
As shown by current observations, there is an enhancement around the M31 direction, but the measurement and physical interpretation involve several uncertainties mainly due to observational limitations.
Here, we summarize these observation limitations and uncertainties, and raise the caveats.
With these limitations, we also discuss the preferred physical properties.

\subsubsection{X-ray and SZ Measurement Uncertainties}
In Section 2.1, we suggested that the enhancement around the M31 direction is unlikely due to a random variation at $6.6 \sigma$, or a Galactic disk variation at $5.7 \sigma$.
However, it is still possible that the MW has contributions to the M31 enhancement.
To have a large-scale ($\approx 20^\circ$) enhancement, the feature is likely to be close to the Sun, so the LB is a possible source for the MW variation.
The LB model used in Section 2.1 is from \citetalias{Miller:2015aa} (the same as \citetalias{Li:2017aa}), which uses the path length derived based on \ion{Na}{1} absorption survey of nearby stars \citep{Lallement:2003aa}.
In the \citetalias{Miller:2015aa} model, the LB has a mean contribution to the observed \ion{O}{7} measurements of 0.83 L.U. with a $1\sigma$ uncertainty of 0.17 L.U., while the contribution to \ion{O}{8}  is negligible.
Toward M31 within $20^\circ$, the LB model suggest a relatively low OVII contribution of 0.72 L.U..
Considering the variation of the LB contributions, the \ion{O}{7} measurement may be reduced by $\approx 0.2$ L.U., which is $<5 \%$ of the \ion{O}{7} enhancement (5.7 L.U.).

The scatter of features is 1.5 L.U. for \ion{O}{7} and 0.5 L.U. for \ion{O}{8} (Fig. \ref{o78}).
It is possible to affect the enhancement at a similar level, which is uncertain due to the limitation of current data (i.e., more sight lines could reduce the statistical uncertainty).
We are conducting a M31 hot halo study (Huang et al. in prep.), which will increase sight lines around M31 within $10^\circ$ (more than 30 sight lines), and improve the current situation.

The uncertainty of the SZ observation is mainly the possible overestimation of the fiducial extraction and the underestimation of the minimal extraction as stated in Section 2.2.
An SZ signal of $y\approx 4\times 10^{-7}$ in the fiducial extraction is about an upper limit of the SZ signal, while the lower limit is $y\approx 2\times 10^{-7}$ extracted in the minimal model.
Although it is possible that the real SZ signal is undetectable ($<2\times 10^{-7}$) because of the $2.5\sigma$ detection in the minimal extraction, the well-behaved radial profiles suggest a more physical reason rather than random variation for both extractions.

One common issue is the coverage of the feature for both SZ and X-ray signals.
There is only a half of area with signals ($b<-20^\circ$), so it is not guaranteed that the upper and the lower halves have similar signals.
In an extreme case, the upper half of $b>-20^\circ$ does not have any enhancement. 
This possibility reduces the angular coverage of the observed features, but will not affect the derived temperature, density, and path length.
However, it will affect the mass estimation as discussed in Section 3.4.3.

\subsubsection{Corrections due to the Multi-Phase Medium}

Although we assume a single-phase model in the toy model, the bridge structure could be multi-phase. 
\citet{Lehner:2020aa} show the covering factor of strong absorption of the UV ion \ion{O}{6} ($\log N > 14.6$) has a peak at $20^\circ$ (Fig. 11 in their paper), which might indicate a transition from the hot gas to the lower-temperature gas (i.e., multi-phase medium).
However, the cool-warm gas contribution associated with the hot bridge is unknown to the observed UV absorption systems.
Therefore, the temperature distribution of the multi-phase medium is also unknown.
We adopt a steady-state cooling assumption ($M\sim T/\Lambda$; roughly a power law with a slope of -1.5; \citealt{Qu:2018aa}) for the warm-hot gas ($\log T > 5$) to investigate the acceptable temperature region.
The \ion{O}{7}, \ion{O}{8}, and SZ measurement is only sensitive to the high-temperature region, so we fix the lower temperature bound to $\log T = 6$.
The maximum temperature cannot be much higher than $\log T=6.6$, otherwise, it predicts less \ion{O}{7} than \ion{O}{8}, while the observed ratio is about $2$.

The multi-phase assumption could significantly reduce the baryon mass and raise the metallicity.
The total mass is proportional to $Y$/$T$ (i.e., electron column density and physical area), so a higher mean temperature gives a lower baryon mass.
At $\log T = 6.3-6.4$, \ion{O}{7} and \ion{O}{8} are at about their emissivity peaks.
If more mass is in the hot phase not contributing to \ion{O}{7} and \ion{O}{8} emission measurements, a higher metallicity is needed to match the observations.
With a high-temperature end of $\log T=6.6$, the total baryon mass will be reduced by $30\%$, and the metallicity will be raised by $80\%$ for both fiducial and minimal SZ models.

In the single-phase models, the derived hot gas masses are about $5.5\times10^{11}~M_\odot$ and $1.9\times10^{11}~M_\odot$ for the fiducial and minimal SZ extraction, respectively (Table \ref{phy_params}).
Considering the multi-phase correction, the masses are reduced to $4.2\times10^{11}~M_\odot$ and $1.4\times10^{11}~M_\odot$.
Also, the metallicity will be raised to $\approx 0.02~Z_\odot$ (fiducial) and $0.12~Z_\odot$ (minimal). 
Although this bias (always higher mass and lower metallicity in the single-phase modeling) is within the uncertainty ($\approx 1 \sigma$), this trade between single-phase and the multi-phase assumptions could play an important role, when accounting for baryons or metals in the hot phase ($\log T>6$).

\subsubsection{Mass and Density and Metallicity}

The modeling based on current data suggests a mass of $1.4-4.2\times 10^{11}~M_\odot$ for the hot bridge.
However, as discussed in Section 3.4.1, the current data has some uncertainties, so here we provide a set of estimations for the lower limit of the hot gas mass.
First, we ignore the SZ signal, which has a higher systematical uncertainty due to the extraction method.
As shown in Fig. \ref{phy_fiducial} and Fig. \ref{phy_min}, the density (proportional to the mass and SZ $y$ signal) and the metallicity are negatively correlated.
Theoretically, there is a solid upper limit of the metallicity of the hot bridge at the Solar metallicity.
Thus, assuming the Solar metallicity, a solid lower limit of the mass is $> 2\times 10^{10}~M_\odot$ (considering \ion{O}{7} and \ion{O}{8} measurement uncertainties), while the preferred mass lower limit is $\approx 8\times 10^{10}~M_\odot$ (with the preferred line measurements) in the multi-phase model.

We also consider the uncertainty of the \ion{O}{7} and \ion{O}{8} measurements by reducing the strength by a factor of 2 (i.e., $\approx 2\sigma$ of the random variation for both ions).
Then, the mass lower limit is reduced to $6\times 10^{10}~M_\odot$.
Because the current coverage of X-ray and SZ measurements is only the lower half ($b<-20^\circ$), the lower limit can be reduced by another factor of 2 (i.e., only the lower half has the bridge structure).
Finally, the lower limit of the bridge mass is about $3-4\times 10^{10}~M_\odot$, so it is still a massive baryonic component in the LG.

The upper limit of the hot gas mass is from the fiducial single-phase model (i.e., $5-6\times 10^{11}~M_\odot$).
Then, the possible range of the mass is between $3\times 10^{10}~M_\odot$ and $6\times 10^{11}~M_\odot$, while it is likely to be higher than $7\times 10^{10}~M_\odot$ ($9\%$ of the LG baryons).
For a mass of $7\times 10^{10}~M_\odot$, the conclusion still holds that the X-ray and SZ enhancement is unlikely to be the M31 hot halo.
If the enhancement is around M31 at 750 kpc, the mass will be increased to $2.8\times 10^{11}~M_\odot$ (a factor of 4), which still breaks the M31 baryonic constraint: a total baryon mass of $1.6-3.2\times 10^{11}~M_\odot$ \citep{Kafle:2018aa}, $\approx 1\times 10^{11}~M_\odot$ in the disk \citep{Tamm:2012aa}, and a massive cool-warm CGM of $>4\times 10^{10}~M_\odot$ \citep{Lehner:2020aa}.

The upper limit of the Solar metallicity also reduces the 1$\sigma$ upper limit of the oxygen mass from $3.4\times 10^8~M_\odot$ to $2.1\times 10^8~M_\odot$. 
Then, the contribution to the missing metals is reduced from $80\%$ to $50\%$ in the LG.

The density is proportional to the mass with the same geometry (cylindrical bridge), which is not sensitive to the \ion{O}{7}, \ion{O}{8}, and SZ $y$ strengths.
Applying a similar estimation as the mass, the range of the density is about $2\times 10^{-4}$ to $10^{-3} \cc$.
This density is similar to the measured hot gas density of $1.2\pm0.9 \times10^{-3} \cc$ of the bridge in the galaxy group HCG 16 (a spiral rich galaxy group; \citealt{OSullivan:2014aa}).

With the multi-phase correction, the preferred metallicity is about 0.02 to 0.12 $Z_\odot$, which is consistent with the X-ray measurements of the HCG 16 intra-group medium \citep{OSullivan:2014aa}.
Here we suggest that it is also possible to be higher (even higher than the Solar metallicity) in the minimal SZ model considering the uncertainty of the $y$ measurement.
The relatively low metallicity implies an intra-group medium origin of the local hot bridge, because the CGM of the MW is expected to have a higher metallicity ($\gtrsim 0.3 ~Z_\odot$; \citealt{Bregman:2018aa}).

\section{Conclusion}
The diffuse hot gas feature projected around M31 is detected in both X-ray emission lines (\ion{O}{7} and \ion{O}{8}) and SZ $y$ signals with a total significance of $7\sigma$ ($4.8\sigma$ of \ion{O}{7}, $4.5\sigma$ of \ion{O}{8}, and $>2.5\sigma$ of SZ).
We rule out the possibilities that this feature is the hot halo around M31, otherwise it exceeds the cosmic baryon fraction.
This hot gas feature cannot be in the MW halo, due to its excessively high thermal pressure ($\sim 100$ times the ambient hot halo), or in the MW disk (too X-ray bright).
A preferred explanation is that this hot gas feature is a structure between the MW and M31.
A cylinder model suggests that the length of this structure is about 400 kpc with a radius of 120 kpc.
Therefore, it is a Local Hot Bridge connecting the hot halos of the MW and M31.
This bridge structure has a temperature of $\approx 2.2\times 10^6$ K, a density of $2\times 10^{-4} - 10^{-3}\cc$, and a metallicity $\approx 0.02-0.12 Z_\odot$.
Such a hot bridge contributes about $\approx 10-50 \%$ ($\approx 0.8-4.2\times 10^{11}~M_\odot$) of the total baryons in the LG, and $\approx 10-50\%$ of the LG missing metals (i.e., oxygen, $\approx 0.4-2.0\times 10^8~M_\odot$).

\acknowledgments
We would like to thank Nicolas Lehner for a thoughtful and constructive referee report.
We also thank Yunyang Li for sharing the MW modeling code of \ion{O}{7} and \ion{O}{8} emission line measurements, and Eric Bell, Hui Li, Monica Valluri, and Cam Pratt for their insightful comments that improved the manuscript.
We gratefully acknowledge support from the NASA Astrophysics Data Analysis Program (ADAP) through award number 80NSSC19K1013. 

\bibliographystyle{apj}
\bibliography{MissingBaryon}

\end{document}